# Exploring Scientists' Working Timetable: Do Scientists Often Work Overtime?

Xianwen Wang[1,2,3]*, Shenmeng Xu[1,2]†, Lian Peng[1,2]†, Zhi Wang[1,2]†, Chuanli Wang[1,2]†, Chunbo Zhang[1,2], Xianbing Wang[4]

[1]WISE Lab, Faculty of Humanities and Social Sciences, Dalian University of Technology, Dalian 116085, China.

[2]School of Public Administration and Law, Dalian University of Technology, Dalian 116085, China.

[3]DUT- Drexel Joint Institute for the Study of Knowledge Visualization and Scientific Discovery, Dalian University of Technology, Dalian 116085, China.

[4]School of Control Science and Engineering, Dalian University of Technology, Dalian 116085, China.

† These authors contributed equally as second authors.

* Corresponding author. Tel.: +86 411 847 072 92-23; fax: +86 411 847 060 82.

Email address: xianwenwang@dlut.edu.cn

**Abstract**: A novel method is proposed to monitor and record scientists' working timetable. We record the downloads information of scientific papers real-timely from Springer round the clock, and try to explore scientists' working habits. As our observation demonstrates, many scientists are still engaged in their research after working hours every day. Many of them work far into the night, even till next morning. In addition, research work also intrudes into their weekends. Different working time patterns are revealed. In the US, overnight work is more prevalent among scientists, while Chinese scientists mostly have busy weekends with their scientific research.

# 1. Introduction

In the science community, achievements always correlates with hard work. Motivated by the pressure of competition from all over the world, most scientists today are living and breathing their work.

When does a scientist begin to work, and when does he/she leave his/her table to go to bed? It's hard to know a scientist's working timetable, and there seems to be no good ways except questionnaire survey (Petkova & Boyadjieva, 1994) or follow-up research (Ledford, 2011). As a result, few previous studies are conducted about this topic.

In this research, we find a novel way to explore scientist's working timetable. Since December 2010, in order to "provide the scientific community with valuable information about how the literature is being used *right now*"

(http://realtime.springer.com/about), Springer has launched a new free analytics tool, namely *realtime.springer.com*. According to the function of this *Realtime* platform (*http://realtime.springer.com/map*), every time one journal article or book chapter is downloaded from Springer, the location (as determined by IP-to-city matching) of downloads is depicted on a world map in real time.

Admittedly, working involves many kinds of behaviors, but it's certain that if someone is downloading literatures, he is definitely working at the table. Thus, to explore scientists' working habits reflected by working hours, we have been conducting this research on the downloads of scientific literatures.

## 2. Data and methods

Our data are retrieved from *http://realtime.springer.com/map*. From the *Realtime* platform, two kinds of data are recorded real-timely for 24 hours a day, which are the downloading location and downloading time. City names are displayed and refreshed momently on the map at exactly the time items are downloaded by someone from this city. In other words, we need to record what cities are downloading items from Springer in every second. There may be tens of downloads all around the world in one single second, so the real-time data need to be recorded at high speed. Another problem is that the webpage of *http://realtime.springer.com/map* sometimes interrupts the information updating process (maybe in every tens of minutes), as Fig. 1 shows. We need to refresh the webpage ever and again to get fresh data. Here we refresh the page every 15 minutes.

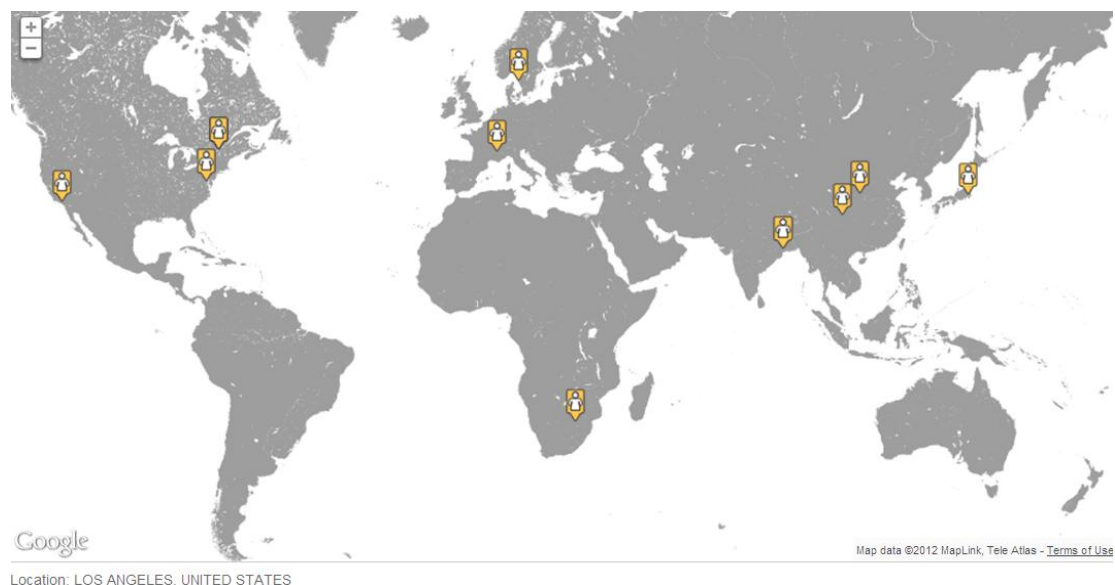

**Fig. 1** Screen shot of the Realtime Download Map (http://realtime.springer.com/map)

We process data of 5 weekdays (April 10, April 11, April 12, April 13, April 16) and 4 weekends (April 14, April 15, April 21 and April 22). We try to monitor and record data in every minute, every second, and even every centisecond completely.

Finally, more than 1,800,000 records are processed and imported into our designed database in SQL Server.

# 3. Results

### 3.1 Daily downloads

Fig. 2 shows the daily downloads on these eight days (GMT). Four weekdays (left four columns) are in blue, while four weekends (right four columns) are in orange. For the weekdays (left four blue columns), the daily downloads are ranged from 180,000 to 21,000. During the weekends, the daily downloads are about 130,000. It can be perceived that there are obvious gaps between the daily downloads of weekdays and weekends.

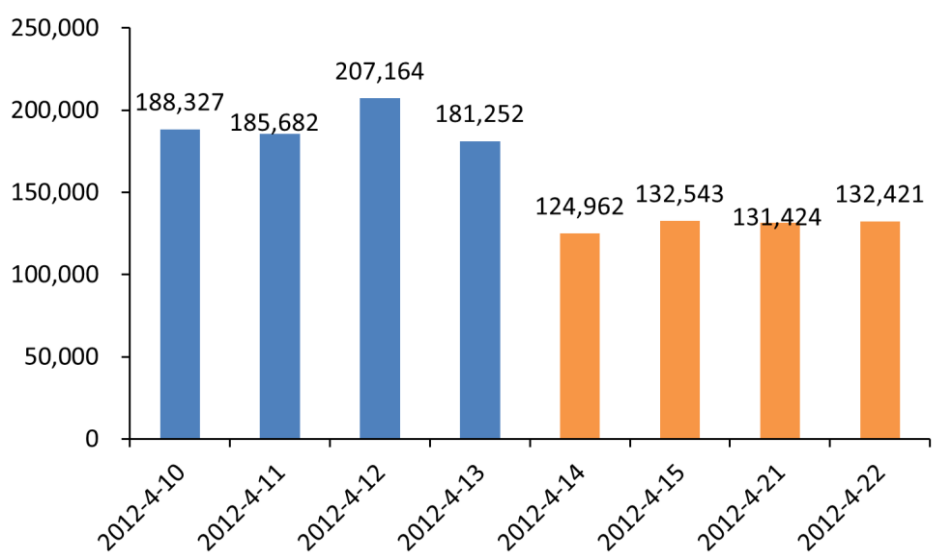

**Fig. 2** Number of downloads on the eight days

### 3.2 Top downloading countries/territories

Gaps of downloads from different countries are also considerable. For example, Table 1 shows the number of downloads of top 20 countries on April 12[th] (GMT). United States has the most downloads of 61,361, accounting for 29.62% of total 207,164 downloads from the whole world. Germany has 31,122 downloads, accounting for 15.02% and ranking the second. Then it follows the Mainland China, who has 19,826 downloads and accounts for 9.57%, ranking the third. In other words, the top 3 countries have 112,309 downloads, and account for 54.21% of all. Downloads from other countries/territories are far behind these 3 countries. In Table 1, the SCI/SSCI publications of these countries are presented. We can see that the downloading counts are to a large extent correlated to the publication counts (SCI/SSCI papers published in 2011) of these countries, except for a minority of countries, such as Italy, Malaysia and Spain.

**Table 1** Top 20 downloading countries/territories on April 12th, 2012

| Rank | Country/Territory | Downloads | Percentage | SCI/SSCI Publications (2011) |
|------|-------------------|-----------|------------|------------------------------|
| 1 | United States | 61,361 | 29.62% | 474,306 |
| 2 | Germany | 31,122 | 15.02% | 115,217 |
| 3 | China | 19,826 | 9.57% | 170,896 |
| 4 | United Kingdom | 8066 | 3.89% | 130,150 |
| 5 | Japan | 6915 | 3.34% | 88,283 |
| 6 | Canada | 6752 | 3.26% | 70,487 |
| 7 | Australia | 6020 | 2.91% | 54,572 |
| 8 | India | 5552 | 2.68% | 50,820 |
| 9 | France | 4880 | 2.36% | 78,327 |
| 10 | South Korea | 4630 | 2.23% | 50,215 |
| 11 | Brazil | 3623 | 1.75% | 39,725 |
| 12 | Netherlands | 3580 | 1.73% | 41,168 |
| 13 | Iran | 3291 | 1.59% | 24,503 |
| 14 | Taiwan | 3247 | 1.57% | 29,592 |
| 15 | Italy | 2938 | 1.42% | 67,361 |
| 16 | Malaysia | 2344 | 1.13% | 8618 |
| 17 | Switzerland | 2221 | 1.07% | 29,393 |
| 18 | Spain | 2119 | 1.02% | 58,905 |
| 19 | Austria | 1905 | 0.92% | 15,667 |
| 20 | Mexico | 1802 | 0.87% | 11,422 |

### 3.3 Hourly downloading in one day

Accordingly, three countries/territories with the most downloads are selected for our analysis, which are the United States of America, Mainland China (Hong Kong, Macao and Taiwan are not included) and Germany.

Here we count the number of downloads hourly for the top 3 countries. A single day is divided into 24 hourly time slices. For example, 0:00:00-0:59:59 is the first slice, and 23:00:00-23:59:59 is the last slice.

Fig. 3 shows the hourly downloading statistics on April 12th, 2012 (GMT). The line with data markers displays the total hourly downloads, and the stacked area chart illustrates the hourly downloads for the top 3 and other countries respectively. As is shown, the total hourly downloads are rather stable in a single day. Generally speaking, during all time slices, the percentages of the top 3 countries are greater than 40% and range from 43.46% to 66.81%. Specifically, during the periods of 0:00-8:00 and 16:00-0:00, the USA dominates the downloading, while during the period of 8:00-16:00, Germany has the most downloads.

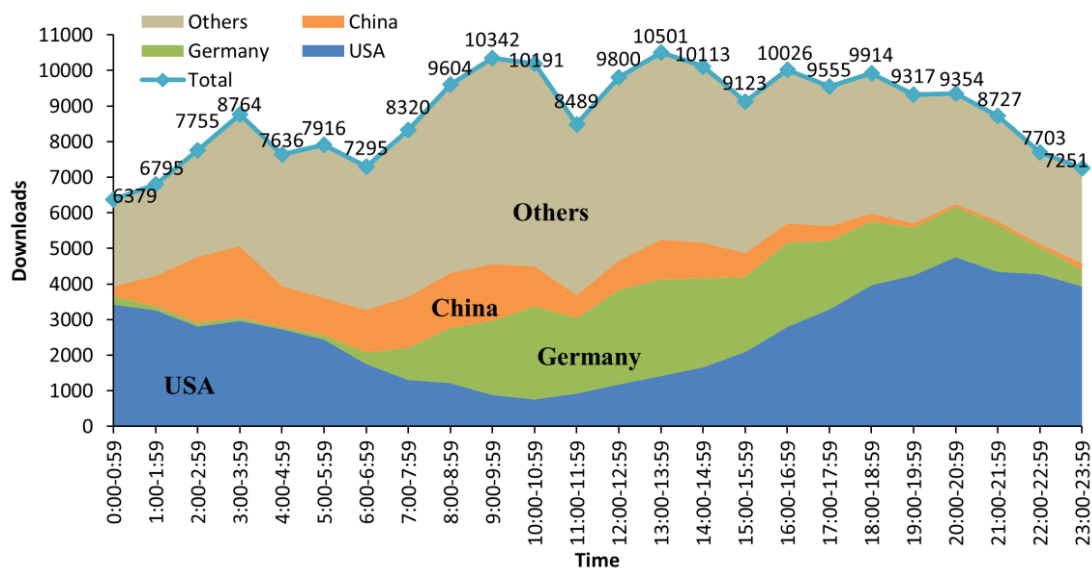

**Fig. 3** Hourly downloading for the top 3 countries on April 12[th], 2012 (GMT)

### 3.4 Comparative analysis of top 3 countries

It is particularly noteworthy that all the time information displayed on the map follows Greenwich Mean Time (GMT), so we need to transform it into local time according to the time zones of the cities.

For countries such as China, Germany, United Kingdom, etc. they have just one time zone. We can transform the downloading Greenwich Mean Time into local time according to the unified time zone offsets. For example, the time zone offset for China is +8.

However, some countries span several time zones, such as the USA, Australia, Canada, etc. The time zone transformation is more complicated for those countries. For example, the United States has 6 time zones. Because the downloading location is determined by IP-to-city matching, and there are 3,208 American cities in the data. Firstly we need to identify the time zone of each city according to the latitude and longitude coordinates, and assign the time zone offset for each city. Then we need to calculate the local time of 561,715 downloading records from these 3,208 cities according to their downloading GMT and time zone offsets.

Here, every 24 hours in a single day are divided into 144 10-minutes slices. As Fig. 4 shows, downloads are counted every 10 minutes. Four weekdays and four weekends are averaged separately. And it needs to be clarified that during the period of 0:00-8:00 on April 11[th], the data of China is extremely abnormal. Further analysis finds that 10-minutes downloads from Tianjin, a city near Beijing, is hundreds of times of the same time on ordinary days. Consequently, to eliminate the error induced by these abnormal data, the four weekdays of China are chosen as April 10, April 12, April 13 and April 16.

If we set time as independent variable (x), the distribution of the number of downloading (f(x)) over time (both weekdays and weekends for each country) is presented in Table 2.

**Table 2** Number of downloading over time for the top 3 countries

| Local time | US weekdays | US weekends | Germany weekdays | Germany weekends | China weekdays | China weekends |
|---|---|---|---|---|---|---|
| 0:00-0:10 | 307.00 | 250.75 | 75.75 | 61.00 | 104.75 | 81.50 |
| 0:10-0:20 | 308.80 | 217.50 | 62.75 | 55.25 | 90.50 | 67.50 |
| 0:20-0:30 | 296.80 | 249.25 | 58.75 | 49.00 | 75.00 | 64.00 |
| 0:30-0:40 | 260.80 | 240.75 | 62.25 | 54.25 | 84.00 | 53.00 |
| 0:40-0:50 | 266.20 | 235.00 | 61.50 | 47.25 | 69.00 | 60.00 |
| 0:50-1:00 | 271.40 | 221.50 | 61.25 | 43.00 | 75.25 | 63.25 |
| 1:00-1:10 | 229.00 | 219.75 | 46.25 | 44.50 | 59.25 | 50.00 |
| … | … | … | … | … | … | … |
| 23:40-23:50 | 305.80 | 272.00 | 88.25 | 73.00 | 126.25 | 71.75 |
| 23:50-0:00 | 292.40 | 240.75 | 80.50 | 87.50 | 140.50 | 79.00 |

The Nonparametric Analysis of Two-Related-Samples Tests is applied. According to the Wilcoxon test results shown in Table 3, the statistical significances of difference are all low at the 0.001 level, which means that these 3 countries have highly different downloading patterns.

**Table 3** Test Statistics[b]

| Test pair | Z | Asymp. Sig. (2-tailed) |
|---|---|---|
| Germany weekdays - US weekdays | -10.410[a] | .000 |
| China weekdays - US weekdays | -10.410[a] | .000 |
| China weekdays - Germany weekdays | -7.601[a] | .000 |
| Germany weekends - US weekends | -10.410[a] | .000 |
| China weekends - US weekends | -10.410[a] | .000 |
| China weekends - Germany weekends | -3.483[a] | .000 |

a  Based on positive ranks.

b  Wilcoxon Signed Ranks Test

Fig. 4 displays the number of downloads according to the local time, thus indicates scientists' research activities during a typical 24-hour day in the top 3 countries. The left three panels display the downloads on the 8 days. Four thin lines illustrate the downloads on four weekdays, while four thick lines display the downloads on four weekends. To better differentiate the downloads on weekdays and weekends, we average the downloads on four weekdays and four weekends separately, as is shown in the right three panels.

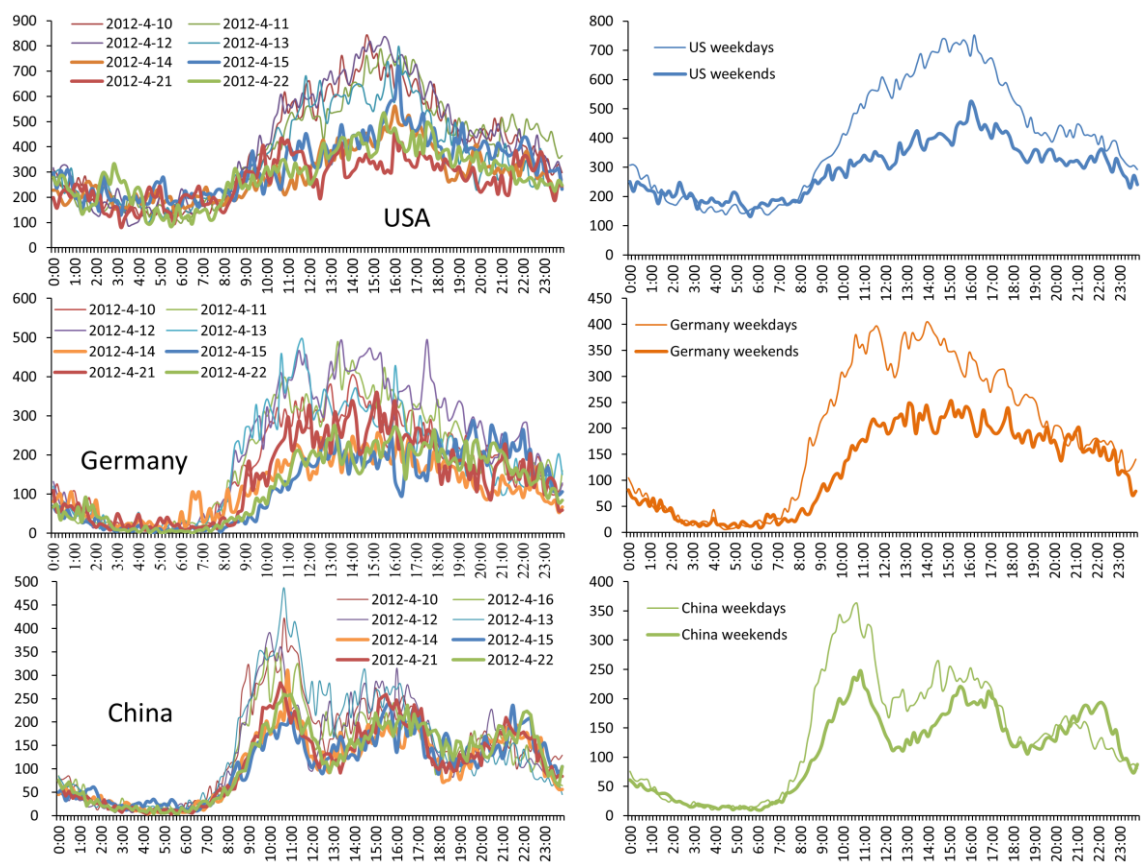

**Fig. 4** Downloads of the top 3 countries during weekdays and weekends

When comparing the downloading curves of weekdays of each country, as the left three panels show, the fluctuations of curves are rather similar. For the weekends, the varying trends of the four curves also remain the same.

As Fig. 4 reveals, all around the world, a fairly large number of scientists are continually engaged in their scientific research after working hours. Many of them work into the early morning hours. In the US, overnight work is particularly prevalent among scientists.

As regards to the whole working timetable, different countries show different features. In the US, the number of downloads keeps rising from about 07:00 in the morning and peaks around 16:00 in the afternoon, which indicates no fixed lunch time during US scientists' day. In Germany, however, the first climax comes around 11:00 in the morning. And then, after a slight decline around 12:00, the lunch time, the second summit comes around 13:00, followed by a fluctuant downward trend afterwards. Nevertheless, in mainland China, the two obvious troughs around 12:00 and 18:00 suggests that scientists tends to put work aside to have a rest during lunch and dinner time, which is probably relevant to the habit and institution that China's dining halls provide food at regular time every day. At about 10:30, 15:30 and 21:00, Chinese scientists' downloads reach the peaks. As a result, a typical Chinese scientist's day is divided into three working periods by these 3 summits and 2 lowebbs.

### 3.5 Comparative analysis of weekdays and weekends

If we define the time period from 0:00 to 08:00 as sleeping time, and from 08:00 to 23:00 as non-sleeping time, here we compare the number of downloads during non-sleeping time at weekdays and weekends based on local time. We can see in Fig. 5 that in these three countries, the number of downloads at weekends account for over 60% of the downloads during weekdays, which fairly demonstrates that a large portion of scientists devote themselves in work at weekends. When we take both Fig. 4 and Fig. 5 into consideration, we find that it's no different for a large portion of Chinese scientists whether it is weekday or weekend. Different from the US all-nighters, Chinese scientists have busy weekends with their scientific research, especially in the afternoon and at night.

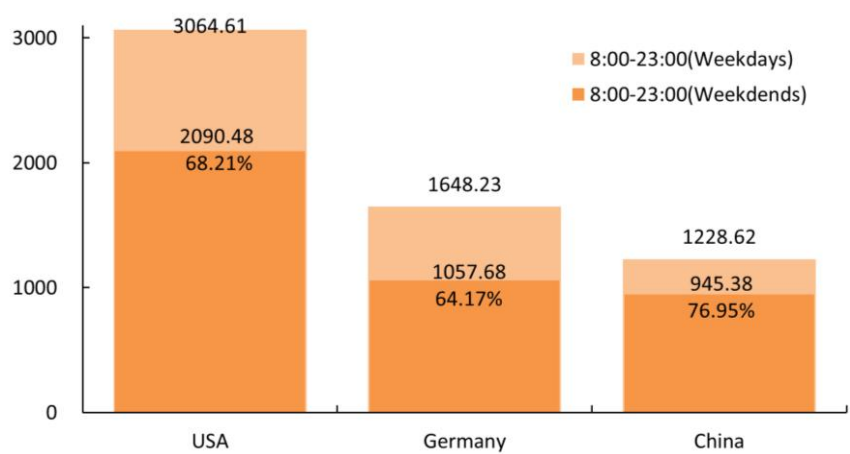

**Fig. 5** Downloads per hour from 08:00 to 23:00 during weekdays and weekends

# 4. Discussions

In recent years, we have seen controversy about whether scientists are sacrificing too much health and family life to achieve more at work (Jacobs & Winslow, 2004; Fox et al., 2011). Scientific achievements are accompanied by intense competition and pressure, which requires a large supply of time and efforts. On the other hand, the demanding assessment from the institution makes the working atmosphere even tenser. Scientists today are spending much more time working than initially intended. They are deprioritizing their hobbies, leisure activities, and regular exercises, which negatively influenced their mental and physical health. Meanwhile, engagement in scientific research after work directly leads to the ambiguity of the boundary between home and office. This investigation on scientists' timetable may in some ways call attention to the unwritten rule of working overtime in academia. As is generally agreed, research is not a sprint but a marathon. Balance in scientists' life is needed.

## Note

During our observation, due to the occasional web accessibility, a very small portion of data is not recorded.   The missing data during the time periods of 0:20-0:29 and 6:40-6:49 on April 10 are replaced by the data between the same time periods on April 17. Similar processing method is applied in 8:30-8:49 and 15:00-15:09 on April 12 (replaced by April 19), 6:00-6:09, 9:30-9:39 and 16:50-16:59 on April 13 (replaced by April 20), 14:50-14:59, 17:10-17:29 and 18:50-19:09 on April 15 (replaced by April 22), 6:30-6.39 on April 21 (replaced by April 14), and 0:00-0:29, 4:30-4:39, 16:50-16:59 and 21:50-21:59 on April 22 (replaced by April 15).

## Acknowledgements

The work was supported by the project of "Social Science Foundation of China" (10CZX011) and the project of "Fundamental Research Funds for the Central Universities" (DUT12RW309).